\pdfoutput=1

\documentclass[12pt,a4paper]{article}

\usepackage{ifthen} 
\newboolean{pdflatex}
\setboolean{pdflatex}{true} 

\newboolean{articletitles}
\setboolean{articletitles}{true} 

\newboolean{uprightparticles}
\setboolean{uprightparticles}{false} 

\usepackage{lineno}  
\usepackage{xspace} 

\usepackage{graphicx}  
\usepackage{color}
\usepackage{colortbl}
\graphicspath{{./figs/}} 

\usepackage{amsmath} 
\usepackage{amssymb}
\usepackage{amsfonts}
\usepackage{upgreek} 

\newcommand*\patchAmsMathEnvironmentForLineno[1]{%
\expandafter\let\csname old#1\expandafter\endcsname\csname #1\endcsname
\expandafter\let\csname oldend#1\expandafter\endcsname\csname
end#1\endcsname
 \renewenvironment{#1}%
   {\linenomath\csname old#1\endcsname}%
   {\csname oldend#1\endcsname\endlinenomath}%
}
\newcommand*\patchBothAmsMathEnvironmentsForLineno[1]{%
  \patchAmsMathEnvironmentForLineno{#1}%
  \patchAmsMathEnvironmentForLineno{#1*}%
}
\AtBeginDocument{%
\patchBothAmsMathEnvironmentsForLineno{equation}%
\patchBothAmsMathEnvironmentsForLineno{align}%
\patchBothAmsMathEnvironmentsForLineno{flalign}%
\patchBothAmsMathEnvironmentsForLineno{alignat}%
\patchBothAmsMathEnvironmentsForLineno{gather}%
\patchBothAmsMathEnvironmentsForLineno{multline}%
}

\usepackage{hyperref}    
\usepackage[all]{hypcap} 

\def\lhcb {\mbox{LHCb}\xspace}
\def\ux85 {\mbox{UX85}\xspace}

\ifthenelse{\boolean{uprightparticles}}%
{

 \def\Ppi         {\ensuremath{\uppi}\xspace}

 \def\PDelta      {\ensuremath{\Delta}\xspace}                 
 \def\PXi      {\ensuremath{\Xi}\xspace}                 
 \def\PLambda      {\ensuremath{\Lambda}\xspace}                 
 \def\PSigma      {\ensuremath{\Sigma}\xspace}                 
 \def\POmega      {\ensuremath{\Omega}\xspace}                 
 \def\PUpsilon      {\ensuremath{\Upsilon}\xspace}                 
 
 \def\PB      {\ensuremath{\mathrm{B}}\xspace}                 
                  
 \def\PD      {\ensuremath{\mathrm{D}}\xspace}

 \def\PK      {\ensuremath{\mathrm{K}}\xspace}

 \def\Pb      {\ensuremath{\mathrm{b}}\xspace}                 
 \def\Pc      {\ensuremath{\mathrm{c}}\xspace}

 \def\Pi      {\ensuremath{\mathrm{i}}\xspace}

 \def\Pp      {\ensuremath{\mathrm{p}}\xspace}

 \def\Ps      {\ensuremath{\mathrm{s}}\xspace}

}
{

 \def\Ppi         {\ensuremath{\pi}\xspace}

 \mathchardef\PDelta="7101
 \mathchardef\PXi="7104
 \mathchardef\PLambda="7103
 \mathchardef\PSigma="7106
 \mathchardef\POmega="710A
 \mathchardef\PUpsilon="7107
                  
 \def\PB      {\ensuremath{B}\xspace}                 
                  
 \def\PD      {\ensuremath{D}\xspace}

 \def\PK      {\ensuremath{K}\xspace}

 \def\Pb      {\ensuremath{b}\xspace}                 
 \def\Pc      {\ensuremath{c}\xspace}

 \def\Pi      {\ensuremath{i}\xspace}

 \def\Pp      {\ensuremath{p}\xspace}

 \def\Ps      {\ensuremath{s}\xspace}

}

\def\squark    {\ensuremath{\Ps}\xspace}

\def\cquark    {\ensuremath{\Pc}\xspace}

\def\bquark    {\ensuremath{\Pb}\xspace}

\def\pion  {\ensuremath{\Ppi}\xspace}

\def\pip   {\ensuremath{\pion^+}\xspace}
\def\pim   {\ensuremath{\pion^-}\xspace}

\def\kaon  {\ensuremath{\PK}\xspace}
  \def\Kbar  {\kern 0.2em\overline{\kern -0.2em \PK}{}\xspace}

\def\Kz    {\ensuremath{\kaon^0}\xspace}
\def\Kzb   {\ensuremath{\Kbar^0}\xspace}
\def\KzKzb {\ensuremath{\Kz \kern -0.16em \Kzb}\xspace}
\def\Kp    {\ensuremath{\kaon^+}\xspace}
\def\Km    {\ensuremath{\kaon^-}\xspace}
\def\Kpm   {\ensuremath{\kaon^\pm}\xspace}

\def\KpKm  {\ensuremath{\Kp \kern -0.16em \Km}\xspace}

\def\Kstarzb {\ensuremath{\Kbar^{*0}}\xspace}

  \def\Dbar    {\kern 0.2em\overline{\kern -0.2em \PD}{}\xspace}
\def\D       {\ensuremath{\PD}\xspace}

\def\Dz      {\ensuremath{\D^0}\xspace}
\def\Dzb     {\ensuremath{\Dbar^0}\xspace}
\def\DzDzb   {\ensuremath{\Dz {\kern -0.16em \Dzb}}\xspace}
\def\Dp      {\ensuremath{\D^+}\xspace}
\def\Dm      {\ensuremath{\D^-}\xspace}

\def\DpDm    {\ensuremath{\Dp {\kern -0.16em \Dm}}\xspace}

\def\Dstarzb {\ensuremath{\Dbar^{*0}}\xspace}

\def\Dstarm  {\ensuremath{\D^{*-}}\xspace}

\def\Ds      {\ensuremath{\D^+_\squark}\xspace}

\def\Dsm     {\ensuremath{\D^-_\squark}\xspace}

\def\Dsmp    {\ensuremath{\D^{\mp}_\squark}\xspace}

\def\B       {\ensuremath{\PB}\xspace}
  \def\Bbar    {\kern 0.18em\overline{\kern -0.18em \PB}{}\xspace}

\def\Bzb     {\ensuremath{\Bbar^0}\xspace}
\def\Bu      {\ensuremath{\B^+}\xspace}

\def\Bp      {\ensuremath{\Bu}\xspace}

\def\Bd      {\ensuremath{\B^0}\xspace}
\def\Bs      {\ensuremath{\B^0_\squark}\xspace}
\def\Bds     {\ensuremath{\B^0_{(\squark)}}\xspace}
\def\Bsb     {\ensuremath{\Bbar^0_\squark}\xspace}

  \def\Y#1S{\ensuremath{\PUpsilon{(#1S)}}\xspace}

\def\proton      {\ensuremath{\Pp}\xspace}
\def\antiproton  {\ensuremath{\overline \proton}\xspace}

\def\Lbar {\ensuremath{\kern 0.1em\overline{\kern -0.1em\PLambda}}\xspace}

\def\Lbbar   {\ensuremath{\Lbar^0_\bquark}\xspace}

\def\to                 {\ensuremath{\rightarrow}\xspace}

\def\CP                {\ensuremath{C\!P}\xspace}

\def\AT#1     {\ensuremath{A_{\mathrm{T}}^{#1}}\xspace}           

\def\C#1      {\ensuremath{\mathcal{C}_{#1}}\xspace}                       
\def\Cp#1     {\ensuremath{\mathcal{C}_{#1}^{'}}\xspace}                    
\def\Ceff#1   {\ensuremath{\mathcal{C}_{#1}^{\mathrm{(eff)}}}\xspace}        
\def\Cpeff#1  {\ensuremath{\mathcal{C}_{#1}^{'\mathrm{(eff)}}}\xspace}       
\def\Ope#1    {\ensuremath{\mathcal{O}_{#1}}\xspace}                       
\def\Opep#1   {\ensuremath{\mathcal{O}_{#1}^{'}}\xspace}                    

\newcommand{\tev}{\ensuremath{\mathrm{\,Te\kern -0.1em V}}\xspace}
\newcommand{\gev}{\ensuremath{\mathrm{\,Ge\kern -0.1em V}}\xspace}
\newcommand{\mev}{\ensuremath{\mathrm{\,Me\kern -0.1em V}}\xspace}
\newcommand{\kev}{\ensuremath{\mathrm{\,ke\kern -0.1em V}}\xspace}
\newcommand{\ev}{\ensuremath{\mathrm{\,e\kern -0.1em V}}\xspace}
\newcommand{\gevc}{\ensuremath{{\mathrm{\,Ge\kern -0.1em V\!/}c}}\xspace}
\newcommand{\mevc}{\ensuremath{{\mathrm{\,Me\kern -0.1em V\!/}c}}\xspace}
\newcommand{\gevcc}{\ensuremath{{\mathrm{\,Ge\kern -0.1em V\!/}c^2}}\xspace}
\newcommand{\gevgevcccc}{\ensuremath{{\mathrm{\,Ge\kern -0.1em V^2\!/}c^4}}\xspace}
\newcommand{\mevcc}{\ensuremath{{\mathrm{\,Me\kern -0.1em V\!/}c^2}}\xspace}

\def\mum  {\ensuremath{\,\upmu\rm m}\xspace}

\def\invfb   {\ensuremath{\mbox{\,fb}^{-1}}\xspace}

\newcommand{\chisq}{\ensuremath{\chi^2}\xspace}

\def\gsim{{~\raise.15em\hbox{$>$}\kern-.85em
          \lower.35em\hbox{$\sim$}~}\xspace}
\def\lsim{{~\raise.15em\hbox{$<$}\kern-.85em
          \lower.35em\hbox{$\sim$}~}\xspace}

\def\pt         {\mbox{$p_{\rm T}$}\xspace}

\def\rad{\ensuremath{\rm \,rad}\xspace}

\def\evtgen     {\mbox{\textsc{EvtGen}}\xspace}
\def\pythia     {\mbox{\textsc{Pythia}}\xspace}

\def\geant      {\mbox{\textsc{Geant4}}\xspace}

\def\tell1  {TELL1\xspace}
\def\ukl1   {UKL1\xspace}

\usepackage{cite} 
\usepackage{mciteplus}

\begin{document}

\renewcommand{\thefootnote}{\fnsymbol{footnote}}
\setcounter{footnote}{1}

\begin{titlepage}
\pagenumbering{roman}

\vspace*{-1.5cm}
\centerline{\large EUROPEAN ORGANIZATION FOR NUCLEAR RESEARCH (CERN)}
\vspace*{1.5cm}
\hspace*{-0.5cm}
\begin{tabular*}{\linewidth}{lc@{\extracolsep{\fill}}r}
\ifthenelse{\boolean{pdflatex}}
{\vspace*{-2.7cm}\mbox{\!\!\!\includegraphics[width=.14\textwidth]{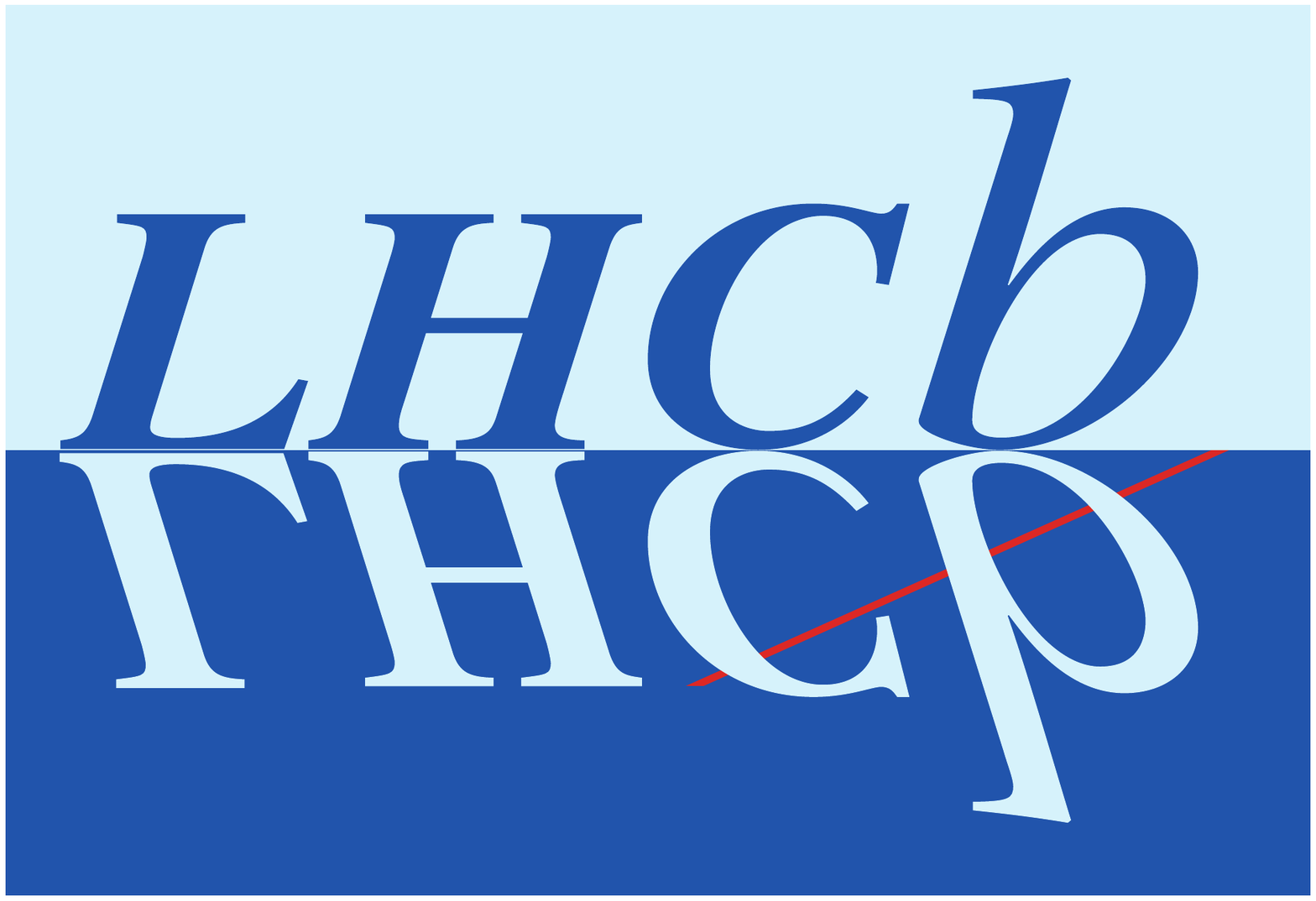}} & &}%
{\vspace*{-1.2cm}\mbox{\!\!\!\includegraphics[width=.12\textwidth]{figs/lhcb-logo.eps}} & &}%
\\
 & & CERN-PH-EP-2012-198 \\  
 & & LHCb-PAPER-2012-018 \\  
 & & August 31, 2012 \\ 
 & & \\
\end{tabular*}

\vspace*{2.0cm}

{\bf\boldmath\huge
\begin{center}
  Observation of the decay $\Bd \to \Dzb\Kp\Km$ and evidence for $\Bs \to \Dzb\Kp\Km$
\end{center}
}

\vspace*{1.0cm}

\begin{center}
The LHCb collaboration\footnote{Authors are listed on the following pages.}
\end{center}\

\vspace{\fill}

\begin{abstract}
  \noindent
  The first observation of the decay $\Bd \to \Dzb K^+K^-$ is reported 
  from an analysis of $0.62 \ \invfb$ of $pp$ collision data collected with the LHCb detector.
  Its branching fraction is measured relative to that of the topologically similar decay $\Bd \to \Dzb \pi^+\pi^-$ to be
  $$
  \frac{
    {\cal B}\left(\Bd \to \Dzb K^+K^-\right)}{
    {\cal B}\left(\Bd \to \Dzb \pi^+\pi^-\right)} = 0.056 \pm 0.011 \pm 0.007 \, ,
  $$
  where the first uncertainty is statistical and the second is systematic.
  The significance of the signal is $5.8\,\sigma$.
  Evidence, with $3.8\,\sigma$ significance, for $\Bs \to \Dzb K^+K^-$ decays is also presented.
  The relative branching fraction is measured to be
  $$
  \frac{
  {\cal B}\left(\Bs \to \Dzb K^+K^-\right)}{
  {\cal B}\left(\Bd \to \Dzb K^+K^-\right)} = 0.90 \pm 0.27 \pm 0.20 \, .
  $$
  These channels are of interest to study the mechanisms behind hadronic $B$ decays, and open new possibilities for \CP violation analyses with larger data sets.
\end{abstract}

\vspace*{1.0cm}

\begin{center}
  Submitted to Phys.\ Rev.\ Lett.
\end{center}

\vspace{\fill}

\end{titlepage}

\newpage
\setcounter{page}{2}
\mbox{~}
\newpage

\centerline{\large\bf LHCb collaboration}
\begin{flushleft}
\small
R.~Aaij$^{38}$, 
C.~Abellan~Beteta$^{33,n}$, 
A.~Adametz$^{11}$, 
B.~Adeva$^{34}$, 
M.~Adinolfi$^{43}$, 
C.~Adrover$^{6}$, 
A.~Affolder$^{49}$, 
Z.~Ajaltouni$^{5}$, 
J.~Albrecht$^{35}$, 
F.~Alessio$^{35}$
% ...

{\footnotesize \it
% ...
$ ^{5}$Clermont Universit\'{e}, Universit\'{e} Blaise Pascal, CNRS/IN2P3, LPC, Clermont-Ferrand, France\\
$ ^{6}$CPPM, Aix-Marseille Universit\'{e}, CNRS/IN2P3, Marseille, France\\
% ...
$ ^{11}$Physikalisches Institut, Ruprecht-Karls-Universit\"{a}t Heidelberg, Heidelberg, Germany\\
% ...
$ ^{33}$Universitat de Barcelona, Barcelona, Spain\\
$ ^{34}$Universidad de Santiago de Compostela, Santiago de Compostela, Spain\\
$ ^{35}$European Organization for Nuclear Research (CERN), Geneva, Switzerland\\
% ...
$ ^{38}$Nikhef National Institute for Subatomic Physics, Amsterdam, The Netherlands\\
% ...
$ ^{43}$H.H. Wills Physics Laboratory, University of Bristol, Bristol, United Kingdom\\
% ...
$ ^{49}$Oliver Lodge Laboratory, University of Liverpool, Liverpool, United Kingdom\\
% ...
$ ^{n}$LIFAELS, La Salle, Universitat Ramon Llull, Barcelona, Spain\\
% ...
}
\end{flushleft}

\cleardoublepage


\renewcommand{\thefootnote}{\arabic{footnote}}
\setcounter{footnote}{0}


\pagestyle{plain} 
\setcounter{page}{1}
\pagenumbering{arabic}


The precise measurement of the angle $\gamma$ of the CKM Unitarity Triangle~\cite{Cabibbo:1963yz,Kobayashi:1973fv} is one of the primary objectives of flavour physics experiments.
Prior to the start of LHC data-taking, the combination of 
measurements with the decay mode $\Bu \to \D\Kp$, where $\D$ denotes a neutral charmed meson that is an admixture of $\Dz$ and $\Dzb$, gave a constraint on $\gamma$ with an uncertainty of around $20^\circ$~\cite{Nakamura:2010zzi}.
Recent results from LHCb on $\Bu \to \D\Kp$~\cite{LHCB-PAPER-2012-001} have helped to reduce this uncertainty, but the use of additional channels to improve further the precision is of great interest. 
The as-yet unobserved decay $\Bs \to \D\phi$ is one of the modes with potential to make a significant impact on the overall determination of $\gamma$~\cite{Gronau:1990ra,Gronau:2004gt,Gronau:2007bh}.
Moreover, a Dalitz plot analysis of $\Bs\to\D\Kp\Km$ can further improve the sensitivity to $\gamma$ due to heightened sensitivity to interference effects, as well as allowing a determination of $\phi_s$, the \CP-violating phase in the $\Bs$--$\Bsb$ system, with minimal theoretical uncertainties~\cite{Nandi:2011uw}.

The first step in the programme towards the measurement of $\gamma$ using the $\Bs\to\D\Kp\Km$ decay is the observation of the channel.
In this Letter the results of a search for neutral \B meson decays to $\Dzb\Kp\Km$ are presented.  
The quantities measured include small contributions from decays to $\Dz \Kp\Km$.
The inclusion of charge conjugate modes is implied throughout.

\begin{figure}[!htb]
  \centering
  \includegraphics[width=0.495\textwidth]{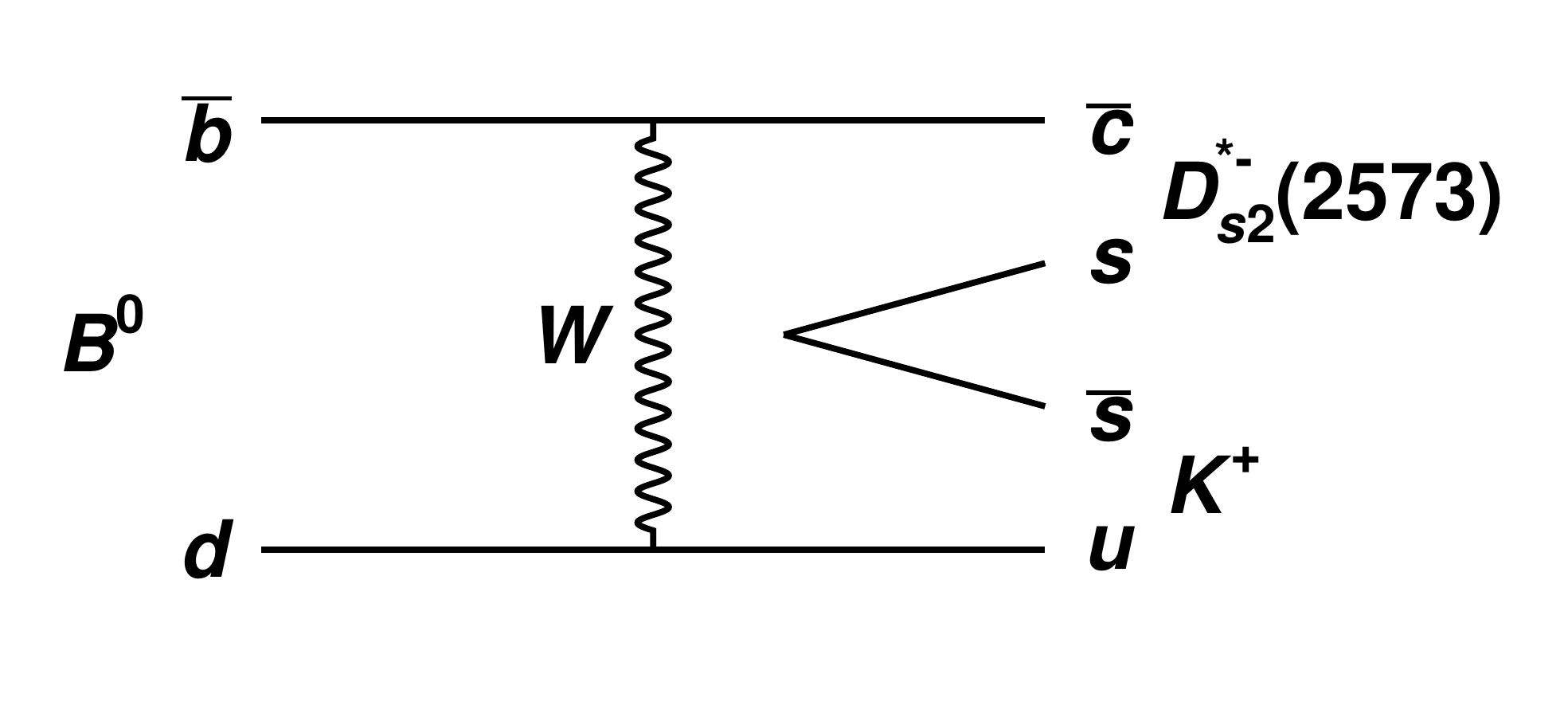}
  \includegraphics[width=0.495\textwidth]{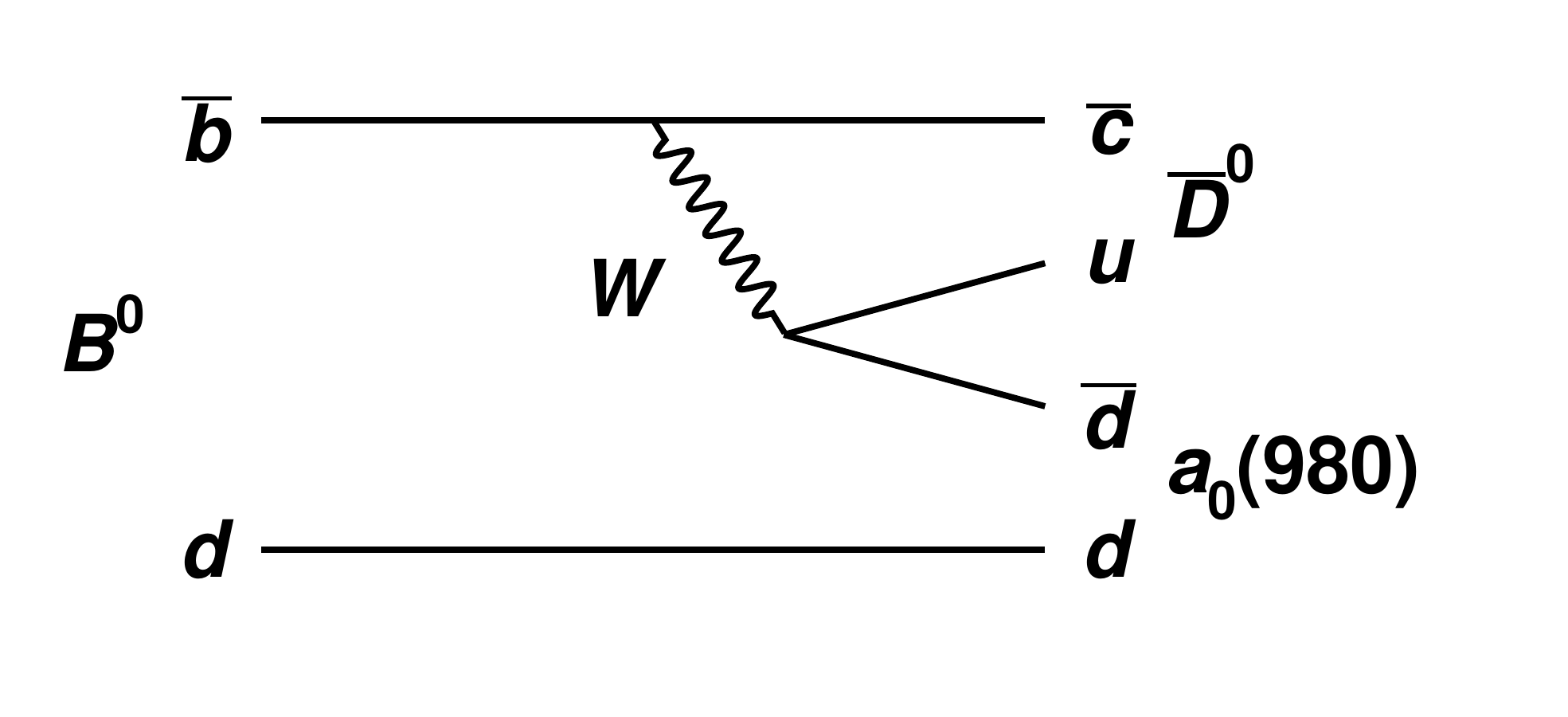}
  \caption{
    Sample decay diagrams that contribute to the $\Bd \to \Dzb\Kp\Km$ final state via (left) $W$-exchange, (right) rescattering from a colour-suppressed decay.
  }
  \label{fig:feynman}
\end{figure}

The analysis uses $0.62 \ \invfb$ of LHC collision data at a centre-of-mass energy of $7 \tev$ collected with the LHCb detector during 2011.
In high energy $pp$ collisions all $b$ hadron species are produced, so both $\Bd$ and $\Bs$ decays are searched for simultaneously.
The decay $\Bd \to \Dzb\Kp\Km$ can be mediated by the decay diagrams shown in Fig.~\ref{fig:feynman}.
These are a $W$-exchange diagram similar to that for the decay $\Bd \to \Dsm\Kp$~\cite{Aubert:2008zi,:2010be} (in this case an excited state that decays to $\Dzb\Km$, such as $D^{*-}_{s2}(2573)$, would be produced), and a colour-suppressed tree diagram producing $\Dzb h^0$, where $h^0$ is a light unflavoured meson such as $a_0(980)$ that subsequently decays to $\Kp\Km$.
Related $B$ decays with $s\bar{s}$ production, $\Bp \to \Dzb \Kp K^{(*)0}$~\cite{Drutskoy:2002ib} and $\Bp \to D_s^{(*)-}\Kp\pip$~\cite{Aubert:2007xma,Wiechczynski:2009rg}, have been measured to have branching fractions of ${\cal O}(10^{-4})$.

The \lhcb detector~\cite{Alves:2008zz} is a single-arm forward
spectrometer covering the \mbox{pseudorapidity} range $2<\eta <5$, designed
for the study of particles containing \bquark or \cquark quarks. The
detector includes a high precision tracking system consisting of a
silicon-strip vertex detector surrounding the $pp$ interaction region,
a large-area silicon-strip detector located upstream of a dipole
magnet with a bending power of about $4{\rm\,Tm}$, and three stations
of silicon-strip detectors and straw drift tubes placed
downstream. The combined tracking system has momentum resolution
$\Delta p/p$ that varies from 0.4\% at 5\gevc to 0.6\% at 100\gevc,
and impact parameter (IP) resolution of 20\mum for tracks with high
transverse momentum ($\pt$). Charged hadrons are identified using two
ring-imaging Cherenkov (RICH) detectors. Photon, electron and hadron
candidates are identified by a calorimeter system consisting of
scintillating-pad and pre-shower detectors, an electromagnetic
calorimeter and a hadronic calorimeter. Muons are identified by a
system composed of alternating layers of iron and multiwire
proportional chambers. The trigger consists of a hardware stage, based
on information from the calorimeter and muon systems, followed by a
software stage which applies a full event reconstruction.
In this analysis, signal candidates are accepted if one of the final state particles created a cluster in the calorimeter with sufficient transverse energy to fire the hardware trigger. 
Events that are triggered at hardware level by the decay products of the other $b$ hadron in the $pp \to b\bar{b}X$ event are also retained.

The software trigger requires characteristic signatures of $b$ hadron decays: at least one track, with high $\pt$
and a large IP with respect to any primary interaction vertex (PV)~\cite{LHCb-PUB-2011-003},
that subsequently forms part of a two-, three- or four-track secondary vertex with a high 
sum of the $\pt$ of the tracks and significant displacement from the PV~\cite{LHCb-PUB-2011-016}.
In the offline analysis, the software trigger decision is required to be due to the candidate signal decay.

Candidates that are consistent with the decay chain $\Bds \to \Dzb \Kp\Km$, $\Dzb \to \Kp\pim$ are selected.
In order to reduce systematic uncertainties in the measurement, the topologically similar decay $\Dzb \pip\pim$, which has previously been well studied~\cite{Kuzmin:2006mw,:2010ip}, is used as a normalisation channel.
The $\Dzb$ candidate invariant mass is required to satisfy $1844 < m_{K\pi} < 1884 \mevcc$.
Tracks are required to be consistent with either the kaon or pion hypothesis, as appropriate, based on particle identification (PID) information from the RICH detectors.
All other selection criteria were tuned on the $\Dzb \pip\pim$ channel.
The large yields available in the normalisation sample allow the selection to be based on data, though the efficiencies are determined using Monte Carlo (MC) simulated events.
For the simulation, $pp$ collisions are generated using
\pythia~6.4~\cite{Sjostrand:2006za} with a specific \lhcb
configuration~\cite{LHCb-PROC-2010-056}.  Decays of hadronic particles
are described by \evtgen~\cite{Lange:2001uf}.
The interaction of the generated particles with the detector and its
response are implemented using the \geant
toolkit~\cite{Allison:2006ve, *Agostinelli:2002hh} as described in
Ref.~\cite{LHCb-PROC-2011-006}.

Selection requirements are applied 
to obtain a clear signal peak in the $\Dzb \pip\pim$ normalisation channel.
The selection includes criteria on the track quality
of the tracks forming the signal candidate, their $p$, $\pt$ and inconsistency with originating from the PV ($\chisq_{\rm IP}$). Requirements are also placed on the corresponding variables for candidate composite particles ($\Dzb$, $\Bds$) together with restrictions on the consistency of the decay fit ($\chisq_{\rm vertex}$), the flight distance significance ($\chisq_{\rm flight}$), and the angle between the momentum vector and the line joining the PV to the $\Bds$ vertex ($\cos \theta_{\rm dir}$)~\cite{LHCB-PAPER-2011-008}.

Further discrimination between signal and background categories is achieved by calculating weights for the remaining $\Dzb\pip\pim$ candidates~\cite{Pivk2005356}. The weights are used by the {\tt NeuroBayes} neural network package~\cite{Feindt2006190} to maximise the separation between categories.
A total of 15 variables are used in the network. They include the $\chisq_{\rm IP}$ of the four candidate tracks, the $\chisq_{\rm IP}$, $\chisq_{\rm vertex}$, $\chisq_{\rm flight}$ and $\cos \theta_{\rm dir}$ of the $\Dzb$ and $\Bds$ candidates,  and the $\Bds$ candidate $\pt$.
Variables describing the $\pt$ asymmetry and track multiplicity in a $1.5 \rad$ cone~\cite{LHCB-PAPER-2012-001} around the $\Bds$ candidate flight direction are also used.
The input quantities to the neural network only depend weakly on the kinematics of the $\Bds$ decay.
A requirement on the network output is imposed that reduces the combinatorial background by an order of magnitude while retaining about 80\,\% of the signal.
No bias is observed by using data driven selection requirements.

To improve the $\Bds$ candidate invariant mass resolution, the four-momenta of the tracks from the $\Dzb$ candidate are adjusted so that their combined invariant mass matches the world average value~\cite{Nakamura:2010zzi}.  
An additional $\Bd$ mass constraint is applied in the calculation of the Dalitz plot coordinates, which are used in the determination of event-by-event efficiencies.  A small fraction ($\sim 5\,\%$ within the mass range described below) of candidates with invariant masses far from the $\Bds$ peak fail the mass constrained fit, and are removed from the analysis.

To remove a large potential background from $\Bd \to \Dstarm(2010)\pip$, candidates in the $\Dzb \pip\pim$ sample are rejected if $m_{\D\pi}$--$m_{\D}$ (for either pion charge) lies within $\pm 2.5 \mevcc$ of the nominal $\Dstarm$--$\Dzb$ mass difference~\cite{Nakamura:2010zzi}.
Candidates in the $\Dzb \Kp\Km$ sample are also rejected if the invariant mass difference calculated under the pion mass hypothesis satisfies the same criterion.
This removes 3.3\,\% of $\Dzb \Kp\Km$ candidates.
Less than 1\,\% of $\Dzb \Kp\Km$ combinations are rejected by requiring that the pion from the $\Dzb$ candidate together with the two kaons do not form an invariant mass in the range $1950$--$1975 \mevcc$, which removes potential background from $\Bs\to\Dsmp\Kpm$ decays.

After all selection requirements are applied, less than 1\,\% of events with at least one candidate also contain a second candidate.
Such multiple candidates are retained and treated the same as other candidates; the associated systematic uncertainty is negligible.

In addition to combinatorial background, candidates may be formed from misidentified or partially reconstructed $\Bds$ decays, or from $\Bds$ decays to identical final states but without intermediate charmed mesons (referred to below as charmless peaking background).  
Contributions from partially reconstructed decays are reduced by requiring the invariant mass of the $\Bds$ candidate to be above $5150 \mevcc$.  
Sources of misidentified backgrounds are investigated using simulation.  Most potential sources are found to have a broad invariant mass distribution, and are absorbed in the combinatorial background shape used in the fit described below.  Backgrounds from $\Lbbar \to \Dzb \antiproton \Kp$ and $\Lbbar \to \Dzb \antiproton \pip$~\cite{LHCb-CONF-2011-036}, $\Bd \to \Dzb \Kp\pim$ and $\Bs \to \Dzb \Km\pip$ decays may, however, give contributions with distinctive shapes and therefore need to be included in the fit.

The contributions from charmless peaking background are investigated using candidates, reconstructed without the $\Dzb$ mass constraint, in sideband regions around the $\Dzb$ mass.
The distributions are fitted with double Gaussian signal and linear background probability density functions (PDFs).
Extrapolating to the $\D$ mass signal region, $773 \pm 30$ ($126 \pm 18$) charmless background decays are expected in the $\Bzb \to \Dzb\pip\pim$ ($\Bzb \to \Dzb\Kp\Km$) distributions.  
No peaking background is observed in the $\Bs$ region.

The signal yields are obtained from unbinned maximum likelihood fits to the $\Dzb\pip\pim$ and $\Dzb\Kp\Km$ invariant mass distributions in the range $5150$--$5600 \mevcc$.
There are 14\,214 $\Dzb\pip\pim$ and 2990 $\Dzb\Kp\Km$ candidates.
The $\Dzb\pip\pim$ fit includes a double Gaussian shape for signal, together with an exponential component for partially reconstructed background, and a PDF for $\Lbbar \to \Dzb X$ decays modelled using a non-parametric function obtained from simulation.  
The $\Dzb\Kp\Km$ fit includes a second double Gaussian component to account for the possible presence of both $\Bd$ and $\Bs$ decays, and peaking background PDFs for $\Lbbar \to \Dzb \antiproton \Kp$, $\Bd \to \Dzb \Kp\pim$ and $\Bs \to \Dzb \Km\pip$, all modelled using non-parametric functions.
The shape of the combinatorial background is essentially linear, but is multiplied by a function that accounts for the fact that candidates with high invariant masses are more likely to fail the $\Bds$ mass constrained fit.

\begin{figure}[!htb]
  \centering
  \includegraphics[width=0.495\textwidth, trim= 25 40 25 50]{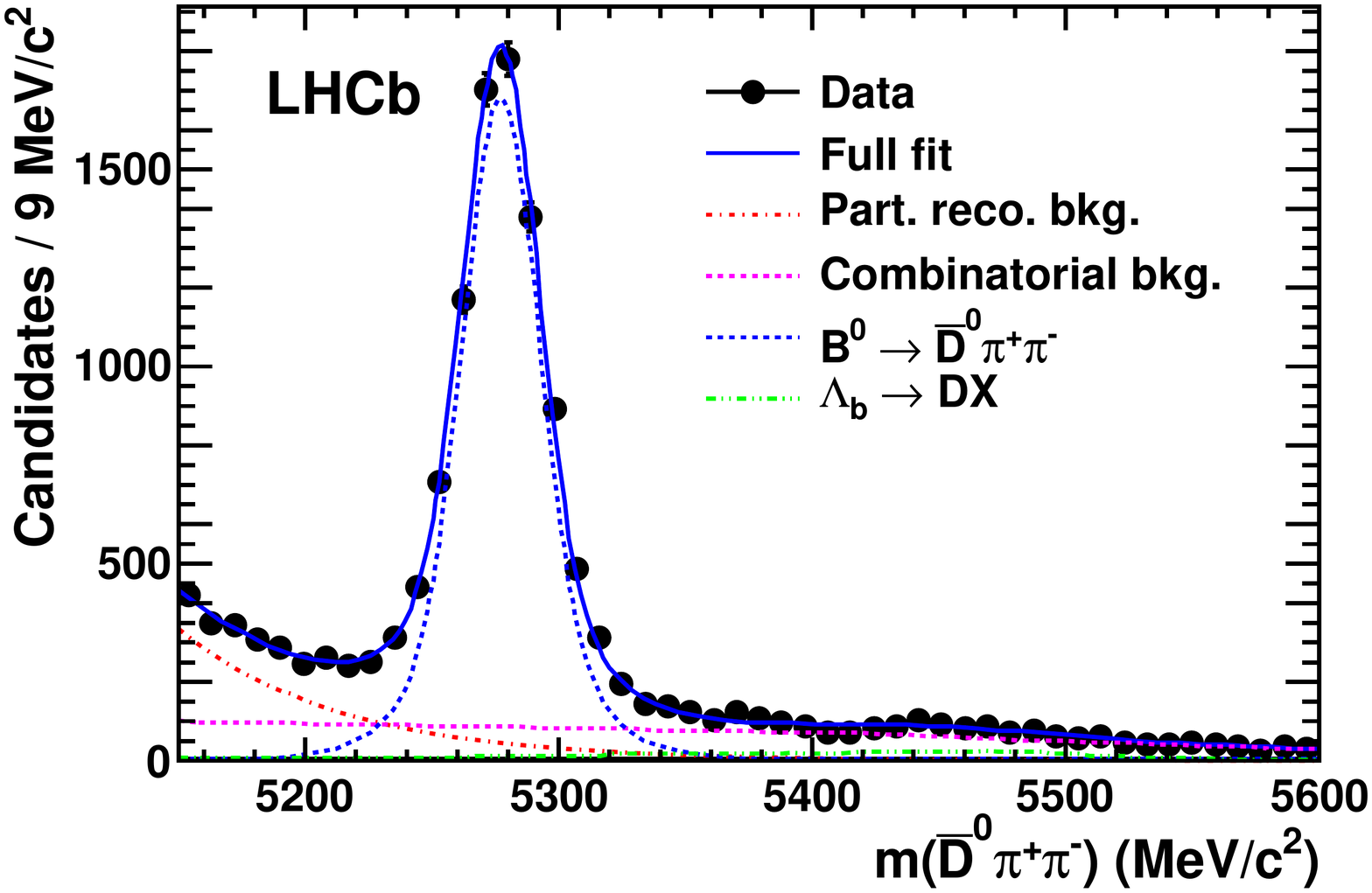}
  \includegraphics[width=0.495\textwidth, trim= 25 40 25 50]{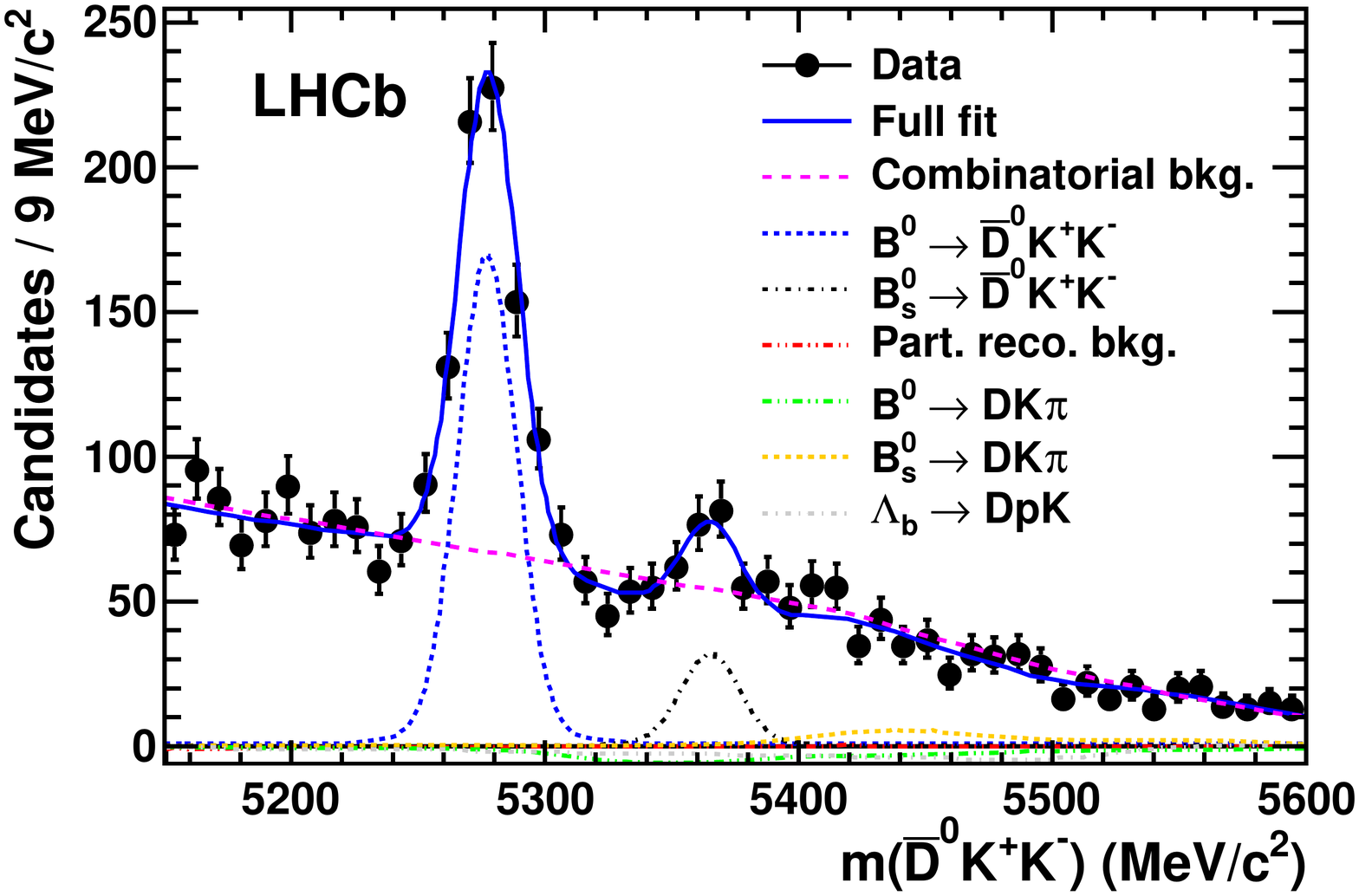}
  \caption{
    Fits to the $\Bds$ candidate invariant mass distributions for the (left) $\Dzb\pip\pim$ and (right) $\Dzb\Kp\Km$ samples. 
    Data points are shown in black, the full fitted PDFs as solid blue lines and the components as detailed in the legends.
    Yields of the partially reconstructed and peaking backgrounds are all small for the $\Dzb\Kp\Km$ sample.
  }
  \label{fig:fits}
\end{figure}

The result of the fit to $\Dzb\pip\pim$ candidates is shown in Fig.~\ref{fig:fits}.
There are nine free parameters in this fit:
the double Gaussian peak position, core width and fraction in the core,
the linear slope of the combinatorial background and the exponential shape parameter of the partially reconstructed background, 
and the yields of the four categories.
The relative width of the broader to the core Gaussian component is constrained within uncertainty to the value obtained in simulation.
The fit yields $8060 \pm 150$ $\Bd \to \Dzb\pip\pim$ decays, including charmless peaking background.

Since the fit to $\Dzb\Kp\Km$ candidates, shown in Fig.~\ref{fig:fits}, has more components, additional constraints are imposed in order to improve the stability of the results.
The parameters of the double Gaussian shapes are constrained to be identical for $\Bd$ and $\Bs$ signals, with an offset in their mean values fixed to the known $\Bd$--$\Bs$ mass difference~\cite{Nakamura:2010zzi}.
The slope of the combinatorial component is constrained to the value obtained in the fit to $\Dzb$ mass sideband events.  
The exponential shape parameter is constrained to the value obtained in the $\Dzb\pip\pim$ fit.
The fit yields $558 \pm 49$ $\Bd \to \Dzb\Kp\Km$ decays, including charmless peaking background, and $104 \pm 29$ $\Bs \to \Dzb\Kp\Km$ decays.
All background yields are consistent with their expectations within uncertainties.  

The ratio of branching fractions is obtained after subtracting the charmless peaking background, and applying event-by-event efficiencies as a function of the Dalitz plot position
\begin{equation}\label{eqn:bfratio}
  \frac{
    {\cal B}\left(\Bd \to \Dzb \Kp\Km\right)}{
    {\cal B}\left(\Bd \to \Dzb \pip\pim\right)} = R(\Bd,\Bd) =
  \frac{
    N^{\rm corr}(\Dzb\Kp\Km) \left( 1 - \frac{N^{\rm peak}(\Dzb\Kp\Km)}{N(\Dzb\Kp\Km)} \right)}{
    N^{\rm corr}(\Dzb\pip\pim) \left( 1 - \frac{N^{\rm peak}(\Dzb\pip\pim)}{N(\Dzb\pip\pim)} \right)} \, ,
\end{equation}
where $N$ is the yield obtained from the fit, $N^{\rm peak}$ is the charmless peaking background contribution, and the efficiency corrected yield
$N^{\rm corr} = \sum_i W_i / \epsilon^{\rm tot}_i$.
Here the index $i$ runs over all candidates in the fit range,
$W_i$ is the signal weight for candidate $i$~\cite{Pivk2005356} from the fit shown in Fig. 2
and $\epsilon^{\rm tot}_i$ is the efficiency for candidate $i$, which depends only on its Dalitz plot position.
The statistical uncertainty on the branching fraction ratio incorporates the effects of the shape parameters that are allowed to vary in the fit, the dilution due to event weighting, and the charmless peaking background subtraction.
Most potential systematic effects cancel in the ratio.  

The PID efficiency is measured using 
a control sample of $\Dstarm \to \Dzb \pim,\,\Dzb \to \Kp\pim$ decays to obtain background-subtracted efficiency tables for kaons and pions as functions of their $p$ and $\pt$~\cite{LHCB-PAPER-2012-002}.
The kinematic properties of the tracks in signal decays are obtained from simulation, allowing the PID efficiency for each event to be obtained from the tables taking into account the correlation between the $p$ and $\pt$ values of the two tracks.
The other contributions to the efficiency (detector acceptance, selection criteria and trigger effects) are determined from simulation, and validated using data.
All are found to be approximately constant across the Dalitz plane, apart from some modulations seen near the kinematic boundaries.

\begin{figure}[htbp]
  \centering
  \includegraphics[width=0.495\textwidth,trim= 25 40 25 50]{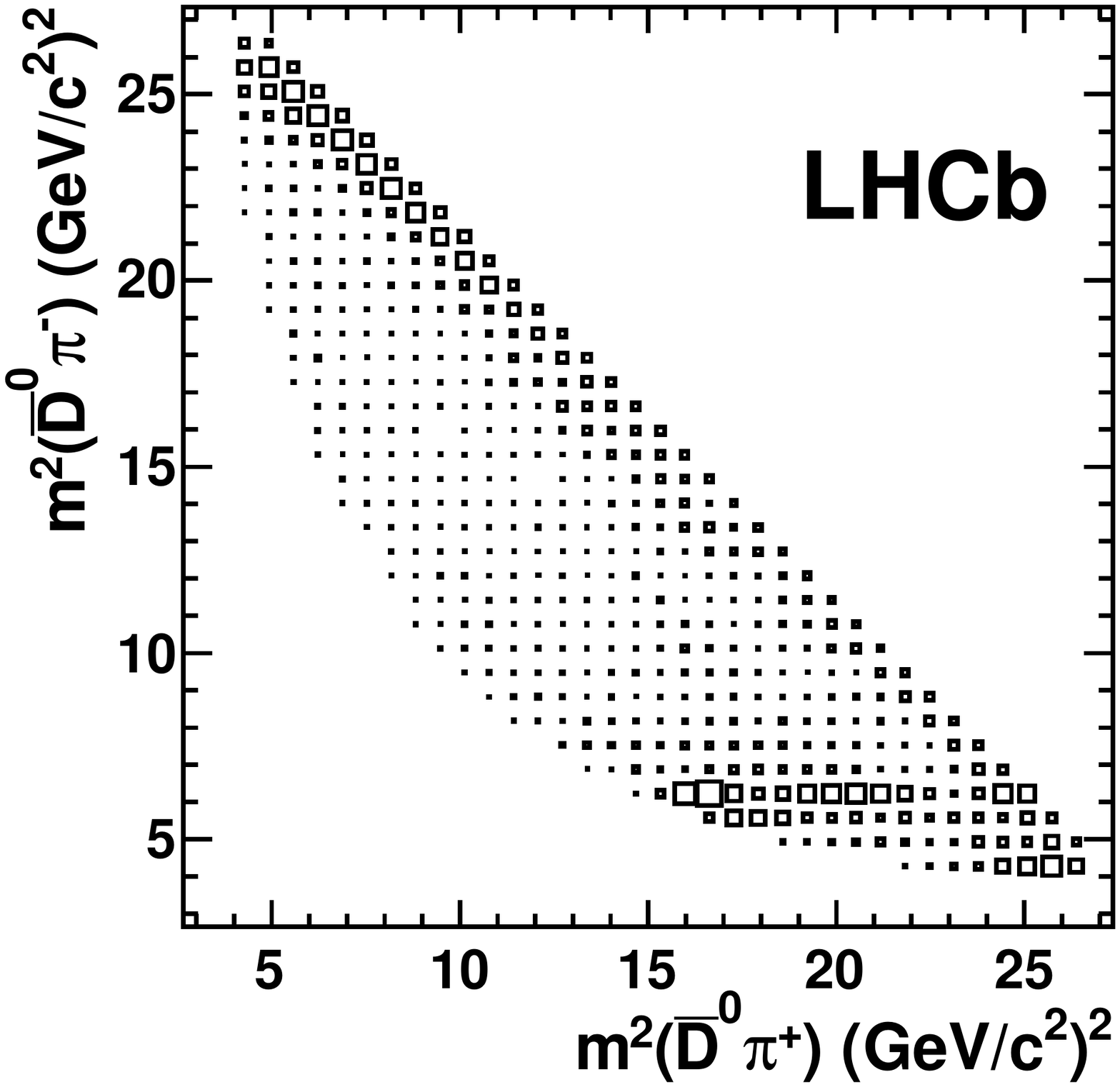}
  \includegraphics[width=0.495\textwidth,trim= 25 40 25 50]{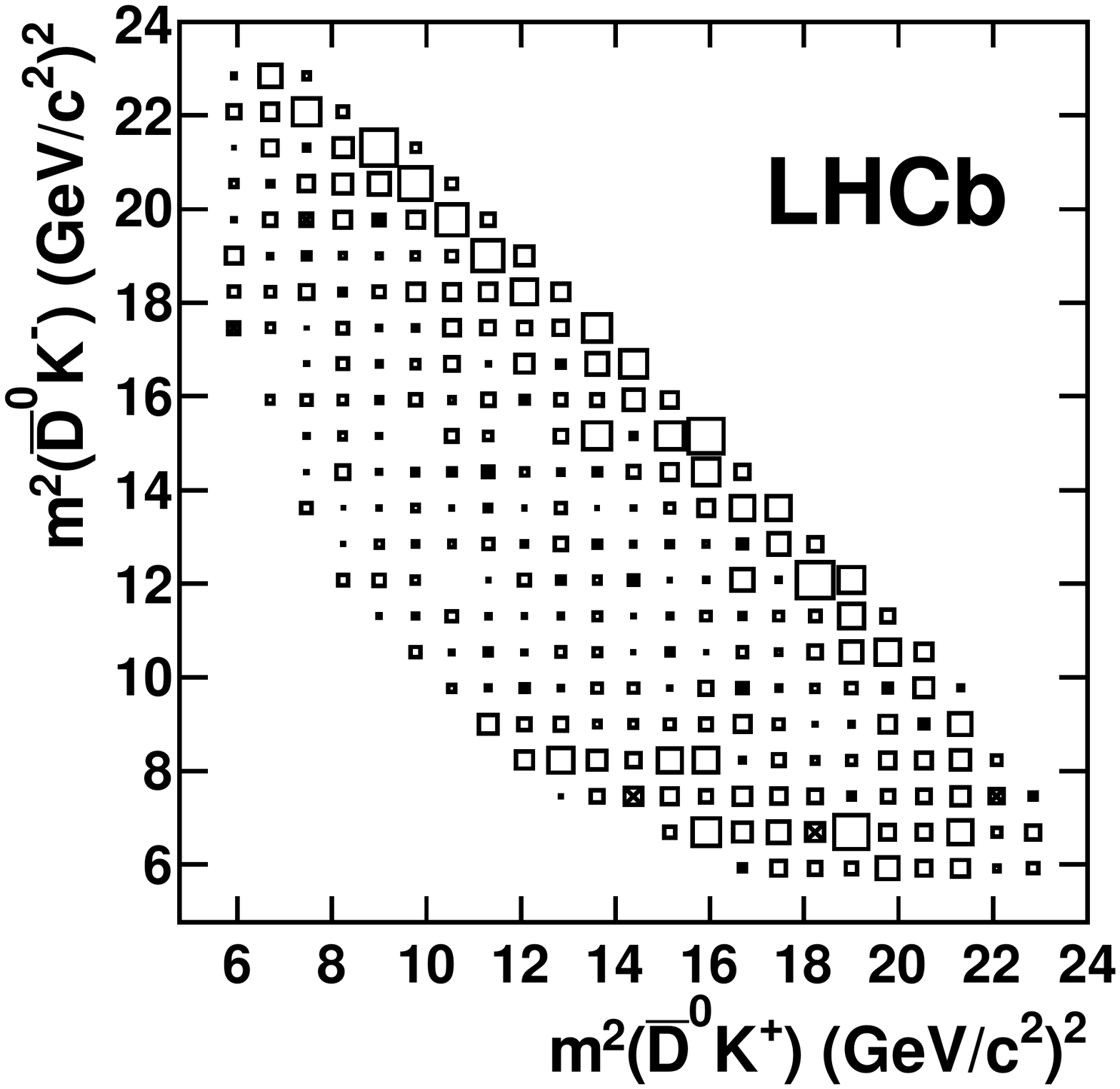}
  \caption{
    Dalitz plot distributions for (left) $\Bd\to\Dzb\pip\pim$ and (right) $\Bd\to\Dzb\Kp\Km$ obtained from the signal weights.
    Note that these distributions contain contributions from charmless peaking backgrounds.
  }
  \label{fig:dalitz}
\end{figure}

The Dalitz plot distributions obtained from the signal weights are shown in Fig.~\ref{fig:dalitz}.
The $\Bd\to\Dzb\pip\pim$ distribution shows contributions from the $\rho^0(770)$ and $f_2(1270)$ resonances (upper diagonal edge of the Dalitz plot)
and from the $\D_2^{*-}(2460)$ state (horizontal band), as expected from previous studies of this decay~\cite{Kuzmin:2006mw,:2010ip}.
The $\Bd\to\Dzb\Kp\Km$ distribution shows a possible contribution from the $\D_{s2}^{*-}(2573)$ resonance, together with an enhancement of events at low $\Kp\Km$ invariant mass (upper diagonal edge).

The branching fraction of the $\Bs$ decay to $\Dzb\Kp\Km$ is measured relative to that of $\Bd$ to the same final state.
Due to the low yield in this decay, an event-by-event efficiency correction is not used.  The ratio of branching fractions is instead determined as
\begin{equation}
  \frac{
    {\cal B}\left(\Bs \to \Dzb \Kp\Km\right)}{
    {\cal B}\left(\Bd \to \Dzb \Kp\Km\right)} = R(\Bs,\Bd) = 
  \left( \frac{f_s}{f_d} \right)^{-1}
  \frac{N(\Bs \to DKK)}{N(\Bd \to DKK) - N^{\rm peak}(\Bd \to DKK)} \, .
\end{equation}
The ratio of fragmentation fractions is 
$f_s/f_d = 0.267\,^{+0.021}_{-0.020}$~\cite{LHCb-PAPER-2011-018}.

Systematic uncertainties are assigned to both branching fraction ratios due to the following sources.
The variation of efficiency across the Dalitz plot may not be correctly modelled in simulation.  
The difference, 6.7\,\%, between the nominal result for $R(\Bd,\Bd)$ and that obtained using Dalitz plot averaged efficiencies is conservatively taken as an estimate of the associated systematic uncertainty.
The fit model is varied by scaling the $\Bs$/$\Bd$ PDF width ratio to account for their different masses, removing components with small yields, adding components for potential background from $\Bs \to \Dstarzb\Kstarzb$ and $\Bs \to \Dstarzb\Kp\Km$, and varying the linear parameter of the combinatorial background PDF within uncertainties from the fit to the $\Dzb$ sidebands used to estimate the charmless peaking background.  
Together these contribute 10.7\,\% (19.9\,\%) to $R(\Bd,\Bd)$ ($R(\Bs,\Bd)$).
An uncertainty of 1.5\,\% is assigned due to the charmless peaking background subtraction procedure.
Possible biases in the determination of the fit parameters are investigated using MC pseudoexperiments, leading to 1.5\,\% (3.4\,\%) uncertainty on the $R(\Bd,\Bd)$ ($R(\Bs,\Bd)$).

In addition, the possible differences in the data/MC ratios of trigger and PID efficiencies between the two channels (both 2.0\,\%) and the effect of the $\Ds$ veto (1.7\,\%) affect only $R(\Bd,\Bd)$.  
The uncertainty on the quantity $f_s/f_d$ (7.9\,\%) affects only $R(\Bs,\Bd)$.
The total systematic uncertainties are obtained as the quadratic sums of all contributions.

A number of cross-checks are performed to test the stability of the result.
The data sample is divided by dipole magnet polarity, data taking period and trigger category. Candidates were divided based upon the hardware trigger decision into three groups;
events in which a particle from the signal decay created a large enough cluster in the calorimeter to fire the trigger, events that were triggered independently 
of the signal decay and those events that were triggered by both the signal decay and the rest of the event.
The neural network and PID requirements are tightened and loosened.  
The PID efficiency is evaluated using the kinematic proprties from $\Dzb\pip\pim$ data instead of from simulation.
The charmless peaking background contribution is determined from the upper and lower $\Dzb$ mass sidebands separately.
All give consistent results.

The significances of the signals are obtained from the changes in likelihood in fits to data with and without signal components, after accounting for systematic uncertainties and for charmless peaking background in $\Bd\to\Dzb\Kp\Km$ only.
They are found to be $5.8\,\sigma$ and $3.8\,\sigma$ for $\Bd\to\Dzb\Kp\Km$ and $\Bs\to\Dzb\Kp\Km$ respectively.

In summary, the decay $\Bd\to\Dzb\Kp\Km$ has been observed for the first time, and its branching fraction relative to that of $\Bd\to\Dzb\pip\pim$ is measured to be
\begin{equation}
  \frac{
    {\cal B}\left(\Bd \to \Dzb K^+K^-\right)}{
    {\cal B}\left(\Bd \to \Dzb \pi^+\pi^-\right)} = 0.056 \pm 0.011 \pm 0.007 \, ,\nonumber
\end{equation}
where the first uncertainty is statistical and the second is systematic.
Using the known value of ${\cal B}\left(\Bd\to\Dzb\pip\pim\right) = (8.4 \pm 0.4 \pm 0.8) \times 10^{-4}$~\cite{Kuzmin:2006mw}, this gives
\begin{equation}
  {\cal B}\left(\Bd\to\Dzb\Kp\Km\right) = (4.7 \pm 0.9 \pm 0.6 \pm 0.5)\times 10^{-5} \,,\nonumber
\end{equation}
where the third uncertainty arises from ${\cal B}\left(B^0 \to \Dzb \pi^+\pi^-\right)$.
Evidence for the $\Bs \to \Dzb K^+K^-$ decay has also been found,
with relative branching fraction
\begin{equation}
\frac{
  {\cal B}\left(\Bs \to \Dzb \Kp\Km\right)}{
  {\cal B}\left(\Bd \to \Dzb \Kp\Km\right)} = 0.90 \pm 0.27 \pm 0.20 \, .\nonumber
\end{equation}
A future study of the Dalitz plot distributions of these decays will provide insight into the dynamics of hadronic $\B$ decays.  In addition, the $\Bs \to \Dzb \Kp\Km$ decay may be used to measure the $\CP$ violating phase $\gamma$.

\section*{Acknowledgements}

\noindent We express our gratitude to our colleagues in the CERN accelerator
departments for the excellent performance of the LHC. We thank the
technical and administrative staff at CERN and at the LHCb institutes,
and acknowledge support from the National Agencies: CAPES, CNPq,
FAPERJ and FINEP (Brazil); CERN; NSFC (China); CNRS/IN2P3 (France);
BMBF, DFG, HGF and MPG (Germany); SFI (Ireland); INFN (Italy); FOM and
NWO (The Netherlands); SCSR (Poland); ANCS (Romania); MinES of Russia and
Rosatom (Russia); MICINN, XuntaGal and GENCAT (Spain); SNSF and SER
(Switzerland); NAS Ukraine (Ukraine); STFC (United Kingdom); NSF
(USA). We also acknowledge the support received from the ERC under FP7
and the Region Auvergne.

\bibliographystyle{LHCb}
\bibliography{main}


\end{document}